# Women, artificial intelligence, and key positions in collaboration networks: Towards a more equal scientific ecosystem

Anahita Hajibabaei[1], Andrea Schiffauerova[1], and Ashkan Ebadi[1,2, *]

[1] Concordia University, Concordia Institute for Information Systems Engineering, Montreal, QC H3G 2W1 Canada
[2] National Research Council Canada, Montreal, QC H3T 2B2, Canada
* Email: ashkan.ebadi@nrc-cnrc.gc.ca

**Abstract**

Scientific collaboration in almost every discipline is mainly driven by the need of sharing knowledge, expertise, and pooled resources. Science is becoming more complex which has encouraged scientists to involve more in collaborative research projects in order to better address the challenges. As a highly interdisciplinary field with a rapidly evolving scientific landscape, artificial intelligence calls for researchers with special profiles covering a diverse set of skills and expertise. Understanding gender aspects of scientific collaboration is of paramount importance, especially in a field such as artificial intelligence that has been attracting large investments. Using social network analysis, natural language processing, and machine learning and focusing on artificial intelligence publications for the period from 2000 to 2019, in this work, we comprehensively investigated the effects of several driving factors on acquiring key positions in scientific collaboration networks through a gender lens. It was found that, regardless of gender, scientific performance in terms of quantity and impact plays a crucial in possessing the "social researcher" in the network. However, subtle differences were observed between female and male researchers in acquiring the "local influencer" role.

**Introduction**

With the increasing growth of the complexity of the scientific projects in terms of the scope and processes, it is crucial for researchers to work collaboratively (Wood and Gray 1991), in diverse and interdisciplinary research teams, to better address the challenges (Bennett and Gadlin 2012). Scientific collaboration in almost every discipline is mainly driven by the need of sharing knowledge, expertise, and pooled resources. The benefits of collaboration are beyond academic publications as it drives innovation and accelerates knowledge dissemination/creation leading to transformative research (Fox and Faver 1984). As a result, it could positively affect research productivity and impact (Ebadi and Schiffauerova 2016a; Servia-Rodríguez et al. 2015), thereby facilitating academic career advancement (Petersen 2015), and providing better access to funding sources (Ebadi and Schiffauerova 2015a). Academic collaboration is a win-win relationship, and its beneficial characteristics have encouraged researchers to adopt more collaborative behavior and made it a topic of burgeoning interest in bibliometric studies (Sonnenwald 2007). The co-authorship network, in which nodes indicate authors and the links between a pair of nodes represent co-authorship relationships, is widely used as a meaningful proxy to measure scientific collaboration (e.g., Glänzel 2001; Newman 2004; Savanur and Srikanth 2010). This mapping could properly manifest mutual direct academic interactions and analyze leading factors to achieve fruitful research, albeit capturing only one main aspect of collaborations (Katz and Martin 1997). The co-authorship network is known as a dynamic and evolutionary network that could affect nodes' structural properties and, as a result, the position and impact of researchers (Barabási et al. 2002).

Owing to the growing importance of research collaboration and its beneficial effects on academic success, several studies have examined the collaborative behavior of scientists and investigated the presence of gender-specific patterns in academic research. Understanding gender aspects of scientific collaboration is of paramount importance (Hajibabaei et al. 2022) since evidence demonstrates the pervasiveness of gender inequality in scholastic activities (Huang et al. 2020; Nelson and Rogers 2003). Several studies reported that men and women exhibit different behavior in collaboration practices, and male scientists are more likely to adopt effective collaborative behavior that could be presumed to lead to higher scientific productivity and impact (Jadidi et al. 2018; Sonnert and Holton 1995). Abramo et al. (2019), for instance, performed a longitudinal analysis and examined the impact of different collaborative patterns on research performance. They found a stronger tendency towards international collaborations among top scientists, which might earn them higher research output, visibility, and reputation. Moreover, it has been shown that women are lagging behind men in international research, and they are generally less active at the international level (Abramo et al. 2013; Larivière et al. 2013; Uhly et al. 2015). Female scientists' collaboration networks contain more weak ties while males tend to have more long-lasting and strong ties (Jadidi et al. 2018), which are generally associated with high-impact research (Petersen 2015). There is also evidence suggesting that scholars favor collaborators of the same gender, identified as the gender homophily effect (Holman and Morandin 2019; Jadidi et al. 2018). Since female scientists are underrepresented in primary disciplines (Hamrick 2019; Holman et al. 2018), the gender homophily effect could create some disadvantages for them, such as less academic recognition, limited access to resources, collaborators, and funding opportunities (Etzkowitz et al. 2000; van den Brink and Benschop 2013). The combination of the aforementioned factors could isolate women, specifically in male-dominant fields, and result in gender productivity gap in academia, i.e., male scholars often outperform females in research activities (Astegiano et al. 2019).

The above evidence demonstrates how different collaboration strategies could affect scholars' academic performance. Several studies applied social network analysis (SNA) to further explore collaborative mechanisms in co-authorship networks and analyze the role of structural network positions in research activities (e.g., Ebadi and Schiffauerova 2015b; Eslami et al. 2013). Network centrality metrics have been utilized extensively by scholars to measure an individual's relative importance and their role within the network. For example, Abbasi et al. (2011) used four centrality indicators, i.e., closeness centrality, degree centrality, betweenness centrality, and eigenvector centrality, and concluded that researchers occupying central positions within the network are often the most influential and critical ones. They also found that some properties, such as having a high number of distinct collaborators, possessing central network positions, and being in close proximity to other authors, are inextricably linked to producing high-impact research. In a similar vein, Uddin et al. (2013) showed that authors engaged in extensive networks, measured by high average degree centrality, with brokerage roles, measured by high betweenness centrality, could take advantage of their positions to publish highly-cited scientific works. The significant positive impact of authors' brokerage role on academic performance is also confirmed in the literature, as brokers can play pivotal roles in the network by tapping diverse resources and facilitating knowledge transmission between unconnected groups (e.g., Abbasi et al. 2012; Ebadi and Schiffauerova 2015b; Gonzalez-Brambila et al. 2013).

Although prior research emphasized the impact of network structural characteristics of researchers on their performance and importance, few studies explored differences between men and women in their positioning and influence in the co-authorship network. In a comprehensive study of computer scientists' collaborative behavior, Jadidi et al. (2018) found that women tend to have smaller and tightly clustered networks and possess fewer brokerage roles compared to their male counterparts. They also demonstrated that while there are no gender differences in collaboration practices among the most successful researchers, women tend to be less involved in effective and potentially fruitful collaborations. In another study, focusing on gender differences in thirty years of collaborations among life science inventors, Whittington (2018) reported that while there are no dissimilarities between the two genders from the network reachability perspective, men are dominant in brokerage positions and benefit more from their strategic roles. Similarly, other studies also confirmed the women's under-representation in critical/influential academic network positions (e.g., Ebadi and Schiffauerova 2016b; Karimi et al. 2019).

Important researchers within the scientific community can are also characterized by their scholarly impact (O'Boyle et al. 2016; Parker et al. 2013), collaborative relationships (Amjad et al. 2016), and their structural positions in the network (Yin et al. 2006). It has been argued that academic research is influenced by a small high-performing pool of scientists as they contribute disproportionately to an ample number of publications with high impact (Xie 2014). Due to this contribution, they can attract more collaborators and resources, and accrue greater recognition, enabling them to parlay their influence into increased career success in academia - a phenomenon dubbed the "Matthew effect" (Merton 1988). At the core of the scientific community, elite researchers may act as ladders for junior ones to reach the pinnacle of academic success and hold bridging positions to control and spread the flow of knowledge (Adegbola 2011). This is closely related to Isaac Newton's famous notion of "standing on the shoulders of giants", denoting that scientific advancement is built on pre-existing work of a few elite scientists whose achievements are strongly inspired by previous work of other eminent researchers. Despite the prominent role of these critical actors in shaping academic networks and mobilizing resources to develop scientific fields (Bonds 2011; Serenko et al. 2011), little is known about which characteristics could make them critical.

Moreover, it is vital to identify influential researchers in disciplines such as artificial intelligence (AI) as a highly interdisciplinary and evolutionary field that faces growing demand for experts to solve unprecedented and challenging global issues (Gagné 2019). AI has dramatically revolutionized every aspect of human life, from impacting critical processes in medical science to intelligent environments and decision-making processes (Makridakis 2017; WIPO 2019). Due to its complex and evolving nature, AI requires knowledge and skills from diverse scientific areas. Thus, effective knowledge diffusion can enhance knowledge and expertise sharing among AI scientists and assist them to overcome common challenges in addressing complex real-world issues. Like many disciplines, it is claimed that AI is also led by a small group of distinguished scientists who are mainly men and form tight clusters together (Yuan et al. 2020), which makes it essential to examine the characteristics of influential AI researchers and explore gender-related patterns in their collaboration network.

All of these spurred us to analyze the collaboration patterns of AI researchers through a gender lens and explore the influence of various factors at the individual level of researchers on achieving strategic and central positions in their surrounding scientific collaboration network. Till far, most previous works focused on the impact of structural network positions on academic collaborative activities and the research performance of scientists. In contrast, this study aims to extend the current literature in reverse order by investigating the effects of driving factors on acquiring key positions in scientific collaboration networks and explaining any possible gender differences. To be more specific, we first considered several author-specific characteristics as independent factors, including but not limited to performance-related metrics, seniority level, and collaborative behavior. We then computed complementary network structure measures, i.e., degree centrality, closeness centrality, and betweenness centrality, as dependent variables and main proxies to measure the impact and importance of researchers. Next, we leveraged machine learning (ML) techniques to identify the most influential researchers whose presence is essential for knowledge and innovation propagation over the network. Finally, using the model interpretation method, we determined the impact of several influencing factors on possessing highly central positions and examine if such factors differ among female and male AI scientists.

This work contributes to the existing research at least in the following ways: (1) it combines SNA and advanced ML techniques and uses a multi-dimensional feature vector at the author level that covers multiple characteristics of scientific activities to identify what factors may lead male and female AI researchers to strategic positions, (2) it investigates the profile of highly central scientists in the AI scientific ecosystem, as an example of a highly interdisciplinary, fast-evolving, and collaborative field, (3) it explores gender differences in the collaborative behavior and characteristics of the most influential AI scientists.

The rest of the paper proceeds as follows: Section 2 presents the data and methods used in this work; Section 3 presents the main findings of the study; We discuss the findings and conclude the paper in Section 4, and finally, our research limitations and potential future research directions are presented in Section 5.

**Data and Methods**

**Data**

In this work, we utilized the same dataset as in our prior study (Hajibabaei et al. 2022). Data on AI-related articles published between 2000 and 2019 were collected from Elsevier's Scopus. We used the following query to extract target publications: ("artificial intelligence" OR "machine learning" OR "deep learning"). Additionally, we extracted journals' ranking information from SCImago and added this information to the database. To detect the gender of AI researchers, we utilized machine learning techniques and natural language processing to develop an automatic gender assignment model. The model was trained on a massive gender-labeled dataset of names to infer the gender using a set of primary features such as full name, affiliation, and country of origin. As a result of the gender assignment model, each author was categorized into female (F), male (M), and unisex/unknown (U). We excluded unisex/unknown authors. To assess the accuracy of our gender determination method, we performed a manual investigation on a random sample of 1000 authors, and the relative accuracy of the gender assignment algorithm was 96% (94% for females, 98% for males). More detailed information about the gender identification algorithm and process is described in Hajibabaei et al. (2022). In the next step, using SNA techniques, we constructed the co-authorship network for the entire examined period, in which nodes represent AI researchers and links between a pair of nodes indicate co-authorship relationships. The final dataset contains 115,717 authors (84,946 men and 30,771 women) contributed to 40,806 AI publications. Given our research objective of identifying profiles of influential AI researchers, we considered three network structure metrics as dependent variables, calculated several independent variables at the author level, and integrated them into the final dataset. The variables and methods used in this work will be introduced in the following sections in more detail.

**Methods**

*Dependent variables*

Using social network analysis, bibliometric analysis, and machine learning techniques, we aim to explore the effect of driving factors on acquiring superior network positions and influential roles within the AI scientific ecosystem. Occupying various positions within the co-authorship network may affect researchers' role, influence, academic performance, and ability to spread or control knowledge/information. Hence, to capture researchers' importance from different perspectives, we calculated three complementary network metrics, including degree centrality, closeness centrality, and betweenness centrality, at the individual level of researchers. Similar to other studies (e.g., Abbasi et al. 2011; Yin et al. 2006), we assumed that highly central researchers are inherently more impactful than others due to their network structural characteristics. Hence, we classified researchers into two categories: "*core*" and "*peripheral*". To do this, we first calculated centrality measures for all the authors in the dataset. Next, for each of the calculated centrality measures, we associated the top 5% of researchers with the highest centrality values as "*core*" and the rest as "*peripheral*". We defined three binary target variables for each of the centrality measures, utilizing which we seek to capture various roles of AI researchers and explore what factors are associated with assuming highly central/core roles that is belonging to the top 5% of scientists with the highest centrality values. Of note, we considered a normalized version of centrality metrics in this work; thus, all of these metrics take a value between 0 and 1. These dependent variables are briefly described as follows:

*Degree Centrality (dc)*

Degree centrality measures how well connected a researcher is within its local neighborhood by counting the number of distinct collaborators. Simply, the degree centrality of node $i$ is calculated by counting the proportion of node $i$'s direct ties as indicated in Equation (1) (Freeman 1978).

$$dc_i = \frac{degree\ of\ node\ i}{highest\ degree\ in\ the\ network} \qquad (1)$$

In the co-authorship network, researchers with high degree centrality values have a high number of distinct collaborators and may influence their network by acting as a transmitter or a receiver of information. The degree centrality can also reflect communicative activities and the popularity of researchers, so we named these researchers "*social researchers*" since they are more likely to highly collaborate with other researchers.

*Closeness Centrality (cl)*

Closeness centrality estimates the importance of a given researcher based on accessibility and their (network) distance to other researchers. More specifically, this metric is calculated based on the sum of the reciprocal shortest-path distance between nodes in the network (Beauchamp 1965). Hence, the closeness of node $i$ can be expressed as:

$$cl_i = \frac{n-1}{\left(\sum_{j \in n-\{i\}} d(i,j)\right)} \qquad (2)$$

where $d(i,j)$ is the shortest-path distance linking node $i$ and $j$, and $n$ is the total number of nodes connected to node $i$ either directly or indirectly. The $cl_i$ metric takes values from 0 to 1, where 1 indicates that node $i$ is located only a hop away from any other nodes, and the value decreases as the total distance between node $i$ and other nodes increases. Researchers with high closeness centrality are, on average, close to most of the network members and can exchange information faster and efficiently through the network. Such researchers can be identified as "*local influencers*" since, due to their prominent positions, they can exert their influence on at least their local community (Ebadi and Schiffauerova 2015b).

*Betweenness Centrality (bc)*

Betweenness centrality is a common proxy to capture a node's ability to pass and control information flows between communities (Freeman 1978). It measures the extent that a particular node is in-between other pairs of nodes (Borgatti 2005). The betweenness of node $i$ is calculated by:

$$bc_i = \sum_{i \neq j \neq k} \frac{\sigma_{jk}(i)}{\sigma_{jk}} \tag{3}$$

wherein $\sigma_{jk}$ is the number of shortest paths connecting node $j$ with $k$ and $\sigma_{jk}(i)$ is the number of those paths including node $i$. Researchers with high betweenness, known as "*gatekeepers*", have pivotal roles in the network since they can bridge unconnected groups of researchers. This could enable gatekeepers to access non-redundant information and control knowledge flows across the entire network.

*Independent variables*

To determine the key factors associated with being highly central/influential in the AI research community, we calculated several independent variables (features) at the individual level of researchers, capturing various aspects of their scientific activities. In the following, we introduce these variables that are used as input to the machine learning model.

*Number of publications.* The total number of past articles published by each author as a measure of research output.

*Average journal rank.* The average rank of the journals in which the author's articles were published, as a proxy of research impact.

*Average citation counts.* The average citations received by the author's articles, as another proxy of the author's research impact.

*Career age.* It is a proxy to measure the author's academic experience represented by the number of years between the author's first and last publications in our dataset.

*Number of distinct co-authors.* The number of distinct coauthors of the author, an indicator that measures how diverse the collaborative behavior of scholars is.

*Average team size*. To consider the effect of scientific team size, we calculated this measure for each author based on the total number of authors in author's papers divided by the total number of author's papers.

*International collaboration ratio.* To capture the degree of internationalization, we calculated the rate of author's articles that have at least one international co-author, i.e., a co-author from a different country (Abramo et al. 2011).

*Contribution impact.* This metric measures authors' contribution share to their publications. There are several methods to measure an author's contribution. We applied the weighted contribution impact technique, which was used in some previous studies (e.g., Ciaccio et al. 2019;

Shapiro et al. 1994). This method considers the first and last authors as the major contributors, second and next to last authors as the second-largest contributors to the research, and so on. It assigns different weights to co-authors according to their byline positions in each article, i.e., each author's contribution equals *1/p* for a given paper where *p* is the author's position in the list. Thus, the researcher's contribution impact is considered the sum of weighted contributions over their articles.

*Funded publication ratio.* The rate of funded articles per author was measured by the number of publications for which the funding source was acknowledged in the paper over all publications of the given author.

*Multi funding sources ratio.* It was defined as the total number of articles with more than one funding source divided by all articles published by each author.

*Disciplinary profile.* We used the topic modeling (TM) technique to infer disciplinary profiles of AI scientists quantitatively and from the content of their publications. We built several TM models with a various number of topics and evaluated them both quantitatively based on topics quality, assessed by metrics such as perplexity and log-likelihood, and qualitatively by three domain experts. We found 8 as the optimal number of topics. Using topic modeling, we extracted document-topic distributions and then defined each author's research profile based on the average topic distributions over their articles and considered the highest probability topic as the main discipline of the author. As a result, we extracted 8 main research subfields, i.e., natural language processing (NLP), genomics-drug discovery, internet of things (IoT)-energy, decision support systems, computer vision-health informatics, unsupervised learning, machine/deep learning, and cyber security-network.

*Disciplinary diversity.* This metric was calculated using the Shannon entropy index, which measures the degree of interdisciplinarity of authors and quantifies the diversity of authors' research profiles based on the variation of fields represented in their articles. Researchers with high disciplinary diversity have published in more diverse research fields and have balanced contributions to those fields.

*Machine learning model*

Using the described independent variables, we trained a machine learning model to classify "*core*" and "*peripheral*" researchers. Specifically, we applied a supervised classification algorithm trained on labeled data and categorized researchers based on the defined labels/roles. We took each mentioned centrality metric as a dependent variable in turn and formulated the model as a supervised binary classifier predicting whether the given researcher belongs to the top 5% of highly central researchers, i.e., the "*core*" class. Moreover, we partitioned the data by gender and conducted experiments on two separate datasets, for female and male researchers, to examine gender differences. Overall, three classification models were built for the female dataset and three models for the male dataset. Specifically, we built and trained XGBoost (Chen and Guestrin 2016) models, an improved version of the gradient tree boosting algorithm, on the defined datasets to classify researchers. XGBoost has become one of the widely used, high-performing, and popular ML algorithms due to its advantageous characteristics such as efficiency, parallelization, ability to handle missing values, invariance to scaling input variables, and robustness to outliers (Nielsen 2016).

As the first step, the data were stratified and randomly split into 90% training and 10% test sets. The test set remained unseen during the training and tunning phases to perform the final model evaluation. For handling missing values, we used mean imputation for numeric features and mode imputation for categorical features since only three variables, i.e., discipline, disciplinary diversity, and international collaboration ratio, contained null values and the missing percentage was less than 0.3%. In the next step, we normalized input data to the range of [0-1] to ensure that features are comparable and have the same scales while reducing the computational complexity and run time. Additionally, our dataset contained imbalanced classes (5% as "*core*" class and 95% as "*peripheral*") which could result in biased predictions towards the majority class. To address this problem, we adjusted the "*scale_pos_weight*" hyperparameter designed for dealing with imbalanced data by considering greater weight for the minority class. We then validated the ML model and optimized hyperparameters using a repeated stratified 10-fold cross-validation module. In this technique, the training data is evenly divided into 10 folds. In each iteration, 9 folds are considered as the training set and the left-out fold as the validation set. This process is repeated 10 times to use each fold as the validation set once. To prevent the bias and obtain a more robust estimated model performance, we repeated stratified 10-fold cross-validation 3 times wherein the training data was split differently. Then the average scores over all repetitions and all folds were reported.

*Model performance evaluation metrics*

To evaluate the classifiers' performance, we adopted commonly used binary classification performance metrics as follows:

*Recall.* This metric indicates the fraction of positive instances (i.e., "*core*" researchers) correctly predicted by the classifier accounted for actual positive instances.

$$Recall = \frac{True\ Positive\ (TP)}{True\ Positive\ (TP) + False\ Negative\ (FN)} \quad (4)$$

*Precision.* It is the number of positive instances (i.e., "*core*" researchers) correctly detected as positive over all positive predictions.

$$Precision = \frac{True\ Positive\ (TP)}{True\ Positive\ (TP) + False\ Positive\ (FP)} \quad (5)$$

*F1 score.* This metric is calculated based on the weighted harmonic mean of precision and recall.

$$F1\ score = \frac{2(Precission * Recall)}{Precission + Recall} \quad (6)$$

*Driving factors identification*

Since we aimed to identify factors driving researchers' roles in the AI research community, i.e., being influential/core vs. follower/peripheral, we utilized the SHapley Additive exPlanations (SHAP) technique (Lundberg and Lee 2017) to determine the magnitude and direction of the effect of features on the model's output. Specifically, SHAP is a game theory-based approach that quantifies the contribution of each feature on a given prediction and then averages over all observations to estimate the overall feature importance. Since SHAP applies a game-theoretic approach, it demonstrates better consistency, robustness to correlated variables, and capability of revealing hidden relationships compared to other model interpretation techniques. It is also well suitable for tree ensemble models (Lundberg et al. 2020; Lundberg and Lee 2017), e.g., XGBoost. Using this technique, we provided not only model interpretability but the potential characteristics of core AI scientists. The conceptual flow of the analytics pipeline is shown in **Error! Reference source not found.**.

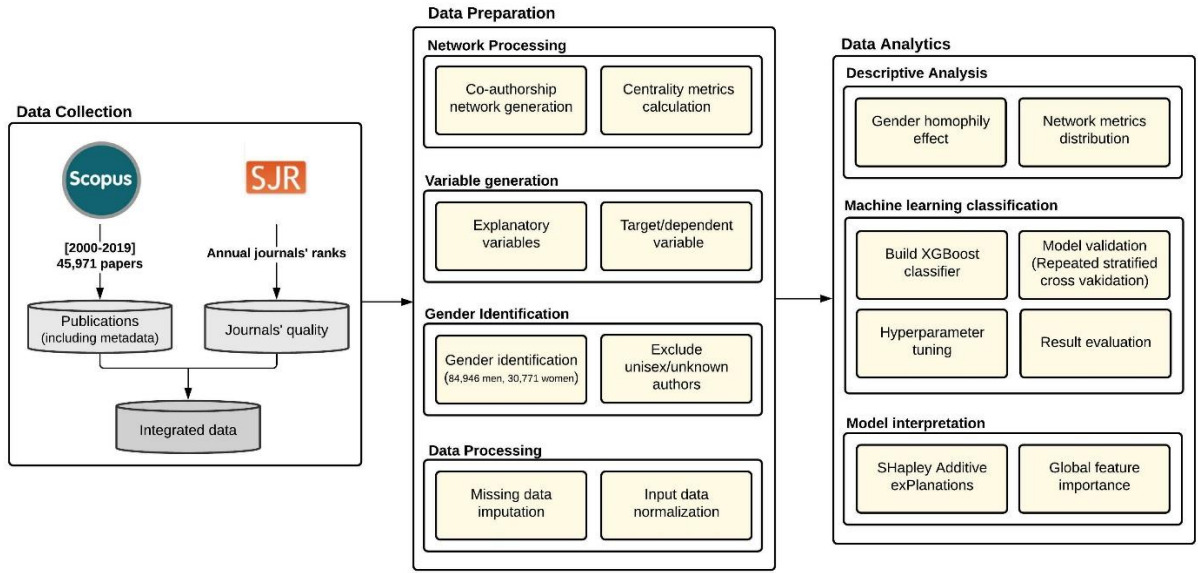

**Figure 1.** The high-level conceptual flow of the analytics pipeline.

## Results

### Descriptive Analysis

The final dataset consisted of 40,806 AI-related publications, published between 2000 and 2019, and 115,717 authors, comprised of 30,771 (27%) women and 84,946 (73%) men. In this section, we discuss descriptive analysis results to provide a better understanding of AI researchers' collaboration patterns and dependent variables' distributions.

*Gender homophily effect*

The gender homophily effect among researchers, i.e., the tendency towards same-gender collaborations, was measured by calculating Yule's Q (odds ratio) index. The gender homophily effect could be strongly influenced by the group size from which researchers can connect with others of their gender. Yule's Q indicator takes into account the effect of different group sizes and is defined as follows (Crossley et al. 2015):

$$Q_i = \frac{IY - EX}{IY + EX} \tag{7}$$

In equation (4), $I$ and $E$ are the numbers of the same and opposite gender collaborators of author $i$, respectively. $Y$ and $X$ are the numbers of collaborators of the opposite and the same gender that author $i$ does not collaborate with. $Q_i$ ranges from −1 (perfect heterophily) to +1 (perfect homophily). Figure 2 depicts the changing trends in AI researchers' collaboration preferences. According to this figure, males show relatively homophilous behavior and predominantly collaborated with other male researchers. However, their scores decreased steadily, indicating they have gradually started to collaborate more with their female counterparts, specifically in recent years. On the other hand, women had more male collaborators in the early years. This is expected as the share of female AI researchers was very small in the beginning years, and their choices were limited. Nevertheless, we can observe that by entering more women into the AI community, female researchers collaborated with researchers of both genders and formed more balanced or gender-neutral research teams, indicated by a decrease in their Yule's Q scores.

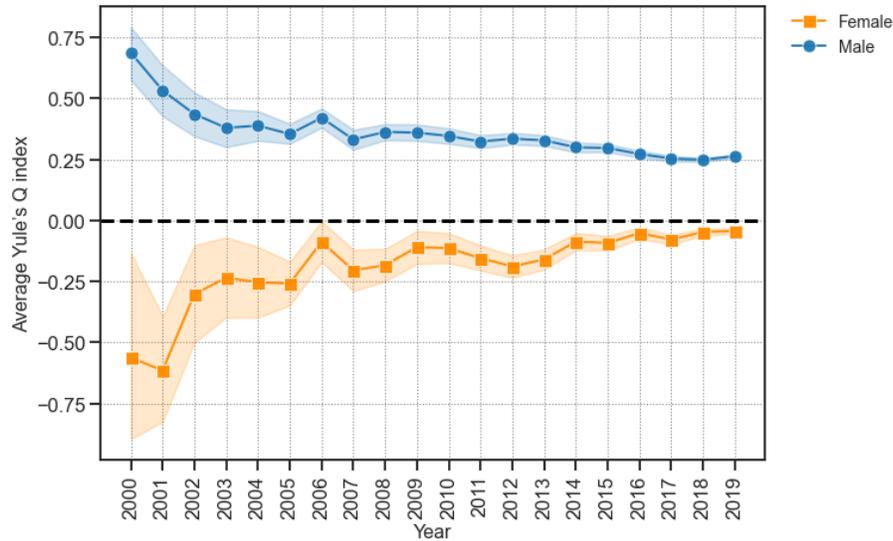

**Figure 2.** Average Yule's Q scores per year for two genders. Shaded areas show a 95% confidence interval. The high variation in the early years is due to the smaller number of researchers in those years compared to the recent years.

*Network metrics distribution*

We next examine the distribution of centrality metrics, i.e., the dependent variables, among two genders shown by box plots in Figure *3*. Centrality values follow right-skewed distribution for both genders, meaning that few researchers occupy highly central positions in the network. Male and female scientists have the same medians and almost the same quartiles for degree and betweenness centrality, but women tend to have higher values in terms of closeness centrality. To perform a statistical comparison between network properties of two genders, we used a one-tailed t-test with 10,000 permutations to identify the significance level without any distributional and independence assumptions. Results revealed that the means of the betweenness and degree centrality for males were significantly higher than those of females (betweenness centrality: $p<0.001$, degree centrality: $p<0.05$). However, female AI scientists show significantly higher closeness centrality on average compared to male scientists ($p<0.001$).

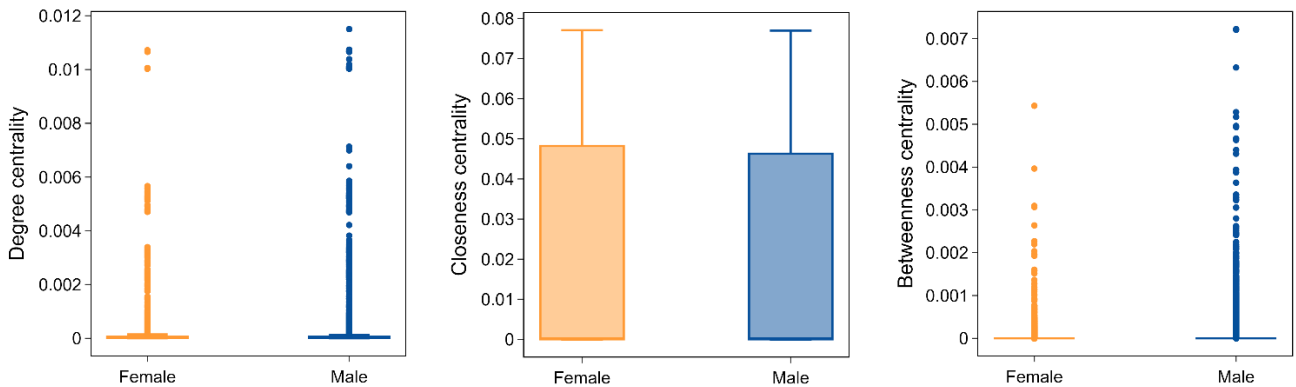

**Figure 3**. Box plots represent the distribution of network centrality metrics for female and male researchers. The box denotes the first (Q1), second (median), and third (Q3) quartiles. Whiskers specify the variation of values outside of the first and third quartiles, and points plotted individually are considered outliers.

**Model performance evaluation**

As explained in section 0, we validated the ML classifiers by applying repeated stratified 10-fold cross-validation. This validation strategy resulted in 30 disjoint validation sets. We used recall, precision, and F1 score (F1) to evaluate our models. Table 1 illustrates the performance of classifiers in detecting the top-5% scientists (core scientists) within different influential role categories (ranked by centrality metrics). As observed, almost all models performed well to predict core scientists in both male and female datasets.

**Table 1.** Model performance validation for predicting core scientists using repeated stratified ten-fold cross-validation. Evaluation metrics were averaged over 30 disjoint validation sets. The table indicates the means and expected variations of performance scores over all repetitions and all folds.

| Prediction of core scientists | Validation sets | Precision | Recall | F1 score |
|---|---|---|---|---|
| Degree Centrality | Female | 0.96 ± 0.01 | 0.93 ± 0.02 | 0.95 ± 0.01 |
|  | Male | 0.94 ± 0.009 | 0.92 ± 0.01 | 0.93 ± 0.006 |
| Closeness Centrality | Female | 0.92 ± 0.02 | 0.82 ± 0.02 | 0.87 ± 0.02 |
|  | Male | 0.87 ± 0.02 | 0.80 ± 0.02 | 0.83 ± 0.01 |
| Betweenness Centrality | Female | 0.72 ± 0.03 | 0.81 ± 0.03 | 0.76 ± 0.02 |
|  | Male | 0.68 ± 0.02 | 0.82 ± 0.01 | 0.75 ± 0.01 |

After validating our models, we further evaluated the performance of classifiers on an unseen test set using the aforementioned metrics to confirm that the results were reliable and generalizable. As indicated in Table 2, we can notice that evaluation results using the test set are almost similar to cross-validation results shown in Table 1, indicating the robustness of our models against unseen data.

**Table 2.** Model performance evaluation using distinct test sets.

| Prediction of core & peripheral scientists | Test sets | Precision | | Recall | | F1 score | |
|---|---|---|---|---|---|---|---|
|  |  | Core | Peripheral | Core | Peripheral | Core | Peripheral |
| Degree Centrality | Female | 0.97 | 1.00 | 0.95 | 1.00 | 0.96 | 1.00 |
|  | Male | 0.93 | 1.00 | 0.93 | 1.00 | 0.93 | 1.00 |
| Closeness Centrality | Female | 0.95 | 0.99 | 0.84 | 1.00 | 0.89 | 0.99 |
|  | Male | 0.88 | 0.99 | 0.81 | 0.99 | 0.84 | 0.99 |
| Betweenness Centrality | Female | 0.71 | 0.99 | 0.84 | 0.98 | 0.77 | 0.99 |
|  | Male | 0.70 | 0.99 | 0.82 | 0.98 | 0.76 | 0.99 |

**Profiles of core AI researchers**

This section addresses our primary research objective, which is to examine the characteristics of the most influential/core AI scientists and explore the differences between male and female researchers. As explained in section 00, using several author-specific characteristics as independent variables, we built several ML classifiers to determine whether the given researcher belongs to the group of the top-5% scientists within each role category, i.e., social researchers, local influencers, and gatekeepers. Then, we utilized the SHAP technique to quantify the importance of features in predictions and investigate the impact of driving factors in possessing core positions within the AI co-authorship network. It should be noted that the SHAP feature importance plot cannot be interpreted as causality and only shows the relationships between features and the predicted target variable (being among the top-5% of highly central researchers). In the following sections, the analysis of characteristics of *social researchers*, *local influencers,* and *gatekeepers* is presented.

*Social researchers*

Figure 4 illustrates feature ranks based on their relative importance in predicting the upper 5% of female and male scientists in terms of their degree centrality values, i.e., social researchers. We excluded the "*Number of distinct co-authors*" and "*Average team size*" features from this analysis as they were highly correlated with the target variable (dc). According to 5, research productivity and scientists' impact, indicated by the number of publications, average journal rank, and average citation counts could positively affect possessing core social positions. This may suggest that productive researchers who publish more high-impact publications can attract more co-authors. One possible reason is that having higher research performance might lead to a higher reputation in the AI community, which could attract other researchers and may provide social researchers with more collaborative opportunities. As seen, a high value of contribution impact, measured based on byline positions, has a great negative impact on the model. In addition, top social researchers have higher disciplinary diversity and balanced contributions to the (sub) fields they are involved in. This is quite expected because having more collaborators could expose them to new knowledge, resulting in publishing in more diverse fields. Moreover, they also benefit from their large collaborative network and access to more (international) co-authors. We can also observe that while career age and funding positively affect possessing core social positions, these factors are the least influencing in the model prediction. Regarding differences between the two genders, the most pronounced difference is that while the number of publications is the most influential factor for men, this factor is less important for women as it ranked fifth. On the other hand, the average journal rank has the greatest influence on acquiring core positions among female social researchers. We can also see the greater magnitude and importance of the disciplinary diversity factor as a proxy for individual interdisciplinarity among female scientists. Lastly, involving in international collaborations has a greater impact on becoming core social researchers for both genders.

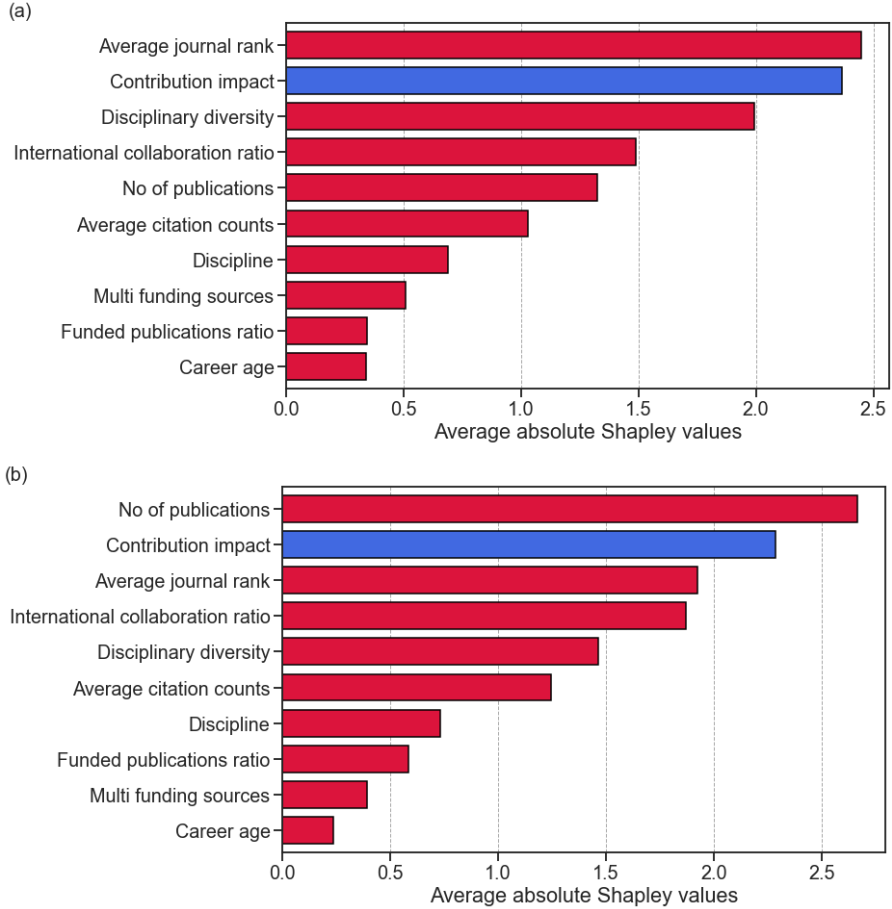

**Figure 4.** These plots indicate the importance of features, in descending order, to predict core **a)** female, and **b)** male social researchers. Features that have positive correlations with being core are shown in red color, otherwise blue. The x-axis denotes by which magnitude each feature contributed to model predictions, which is calculated based on average absolute Shapley values over all observations.

*Local influencers*

Figure 5 illustrates features' importance in classifying core/peripheral AI researchers in terms of their closeness centrality. As observed, average journal rank is the strongest predictor for both genders. The positive effect of average citation counts is also noticeable, even though the number of publications has a relatively small positive impact on possessing core roles among female and male local influencers. A large negative impact of disciplinary diversity can be observed from the feature importance ranking plot. This motivated us to perform further investigation by plotting the probability density distribution of closeness centrality values for researchers involved in less or more than 5 research subfields. It was noticed that these researchers are mostly involved in AI subfields (more than 5 subdisciplines). Involving in more research areas may make researchers unable to have balanced contributions to all those subfields leading to lower disciplinary diversity value (Hajibabaei et al. 2022). Interestingly, we can observe that the discipline is the fourth most informative feature driving the prediction of core scientists for both genders. Hence, we further assessed the role of this factor and found that while core local influencers generally engaged in some specific AI subfields, such as NLP/vision, genomics, and health informatics, they are less involved in research areas such as energy/environment and cybersecurity. In addition, the number of distinct co-authors, average team size, and international collaboration ratio, as proxies to measure the collaborative behavior of scientists, contribute positively to the model prediction. However, there are more commonalities than differences between the two genders. We can see some differences in the collaborative behavior of top local influencers between the two genders. While average team size has a greater effect and importance on the prediction of core female researchers, this factor is less important for males. Other features, namely funded publication ratio, multi funding sources, career age, and the number of publications, are also ranked as the least influencing factors based on their contribution to prediction scientists assuming "*core*" roles.

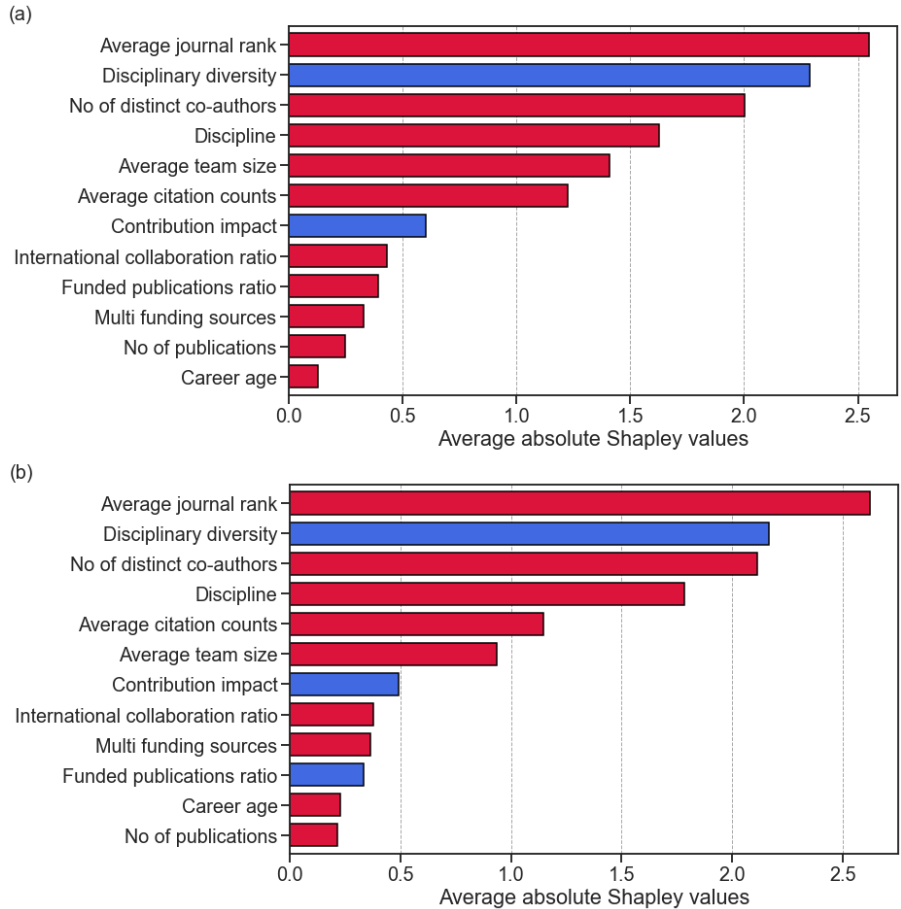

**Figure 5.** These plots indicate the importance of features, in descending order, to predict core **a)** female, and **b)** male local influencers. Features that have positive correlations with being core are shown in red color, otherwise blue. The x-axis denotes by which magnitude each feature contributed to model predictions, which is calculated based on average absolute Shapley values over all observations.

*Gatekeepers*

Lastly, we analyzed the influencing factors in occupying gatekeeper roles, measured by betweenness centrality. As seen in Figure 6, the number of distinct co-authors is the most important driving factor that could positively affect the possibility of occupying core positions among female and male gatekeepers. We can also notice that this feature has a bit higher magnitude, measured by SHAP value, for men than women. The next driving feature is the number of publications, followed by the average team size for both genders. Suggested by these findings, highly productive and collaborative researchers are more likely to occupy highly central positions, maintaining influential brokerage roles within the AI scientific community. Interestingly, the high value of average team size has a negative effect on the prediction of the core class, indicating that top gatekeepers, in general, are involved in smaller teams compared to peripheral ones. In addition, career age as a proxy for seniority level plays an important role in predicting core researchers, and this factor ranked higher for female core scientists. As observed, although high research performance, measured by average journal rank and citation counts, could lead female and male scientists to possess top brokerage positions, these features are a bit more important predictors for core male scientists. We can also see a positive impact for disciplinary diversity, meaning that the most influential gatekeepers generally have more balanced and diverse research profiles. This is fairly expected because these researchers are in superior network positions and could access a variety of knowledge and skill resources by bridging unconnected groups. Another gender difference is that disciplinary diversity is a more important feature for men. Finally, compared to discussed factors, funded publication ratio, multi funding sources, discipline, and contribution impact have small positive impacts on the "*core*" class prediction for both genders.

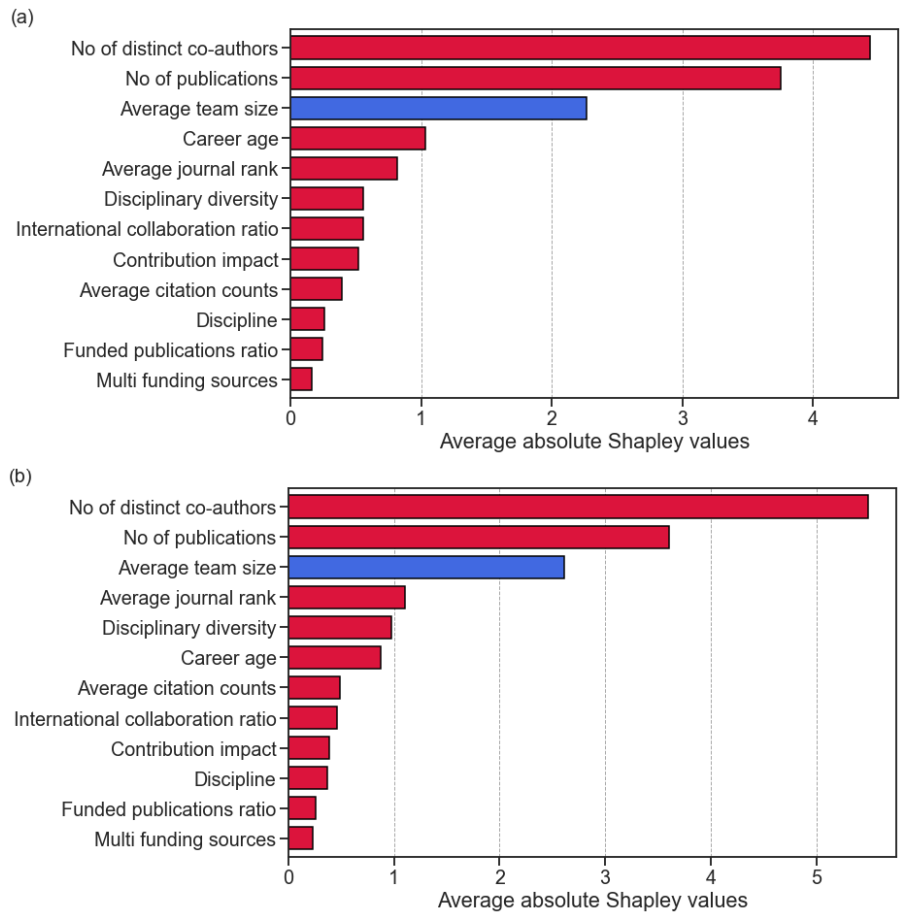

**Figure 6.** These plots indicate the importance of features, in descending order, to predict core **a)** female, and **b)** male gatekeepers. Features that have positive correlations with being core are shown in red color, otherwise blue. The x-axis denotes by which magnitude each feature contributed to model predictions, which is calculated based on average absolute Shapley values over all observations.

**Discussion and conclusion**

As science has become more complex, scientists are inclined to adopt more collaborative behavior, enabling them to benefit from diverse knowledge sources and address perplexing problems. Artificial Intelligence (AI) is a highly interdisciplinary and complex field involving a vast variety of research areas, and its advancement necessitates a diverse set of skills as well as a significant amount of R&D funding. As a key driver of the most current wave of scientific and technological revolution, AI has drastically influenced human lives and ways of thinking, and considerably affected social, industrial and economic activities (Howard 2019). In recent years, AI evolution draws scientists' attention more than before, as indicated by increasing trends in the number of researchers, scientific collaborative activities, and AI publications (AI Index 2019).

Moreover, it is also believed that the AI research and industry community confront rising demand for AI specialists with diverse expertise and face a significant gender gap and lack of gender diversity. This situation could negatively impact innovation activities and even bring unfairness and discrimination risks in AI development (UNESCO 2020). Thus, it is vital to form teams with more qualified and diverse AI experts and create collaboration networks with effective knowledge sharing among actors. In this research, utilizing a combination of social network analysis and machine learning techniques, and a multi-dimensional feature vector at the researcher level that covers multiple characteristics of their scientific activities, we first explored the characteristics of influential/central AI researchers as they can accelerate knowledge/innovation diffusion and form more efficient collaborations. And, we then investigated any possible gender differences in acquiring strategic positions in AI scientific collaboration networks.

According to our findings, regardless of gender, performance metrics, measured by the number of publications, citation counts, journal impact factor, and being involved in more diverse research areas and having a higher degree of internationalization play crucial roles in acquiring network positions with a high degree centrality. At the same time, we observed a stronger impact of publishing papers in more diverse research areas and higher-rank journals on degree centrality among female social researchers, suggesting that they might gain more from their direct and distinct collaborators than male social researchers.

Our results indicated subtle differences between female and male AI researchers when influential researchers are defined based on their number of close collaborators and the degree of reachability (higher closeness centrality). In general, local influencers tend to produce high-impact work and be highly collaborative. Interestingly, we found that discipline is one of the prominent factors that can increase the chance of acquiring influential positions. Local influencers are more likely to be active in disciplines such as NLP, genomics-drug discovery, and computer vision-health informatics, and they, on average, publish in more diverse research areas.

Lastly, it was observed that a high number of distinct co-authors stands out as the leading factor affecting the possibility of possessing brokerage roles. Gatekeepers, i.e., researchers with high betweenness centrality, have a high number of distinct collaborators, but on the other hand, they tend to form smaller research teams. This could imply that it is not necessarily needed to be part of big teams to obtain research excellence. Furthermore, our findings suggest that gatekeepers are the ones who are more likely to have higher seniority levels and scientific performance in terms of both quantity and impact.

In this work, we provided a deeper understanding of the profile of highly central/influential male and female scientists in the AI scientific ecosystem within the 2000 to 2019 period. It was demonstrated that various individual researcher-level factors could contribute differently to occupying different key/strategic network positions in the AI collaboration network. However, some of the notable characteristics of central researchers, regardless of their gender, are their highly collaborative behavior and high research productivity and impact. These central researchers can facilitate and accelerate knowledge flow and exchange across the network and form more efficient and effective scientific collaborations. Moreover, they can assist other researchers in gaining better access to different skills and knowledge resources and, in general, form a more cohesive and efficient collaboration network. Thus, in co-authorship network analysis, it is critical to distinguish between core and peripheral researchers, i.e., individuals who stand out as core players in network structural metrics such as degree, betweenness, and closeness centrality or others who hardly act as intermediaries or nurture overall cohesion and connectivity of the network.

We believe that this work could be an important step in using ML algorithms to explore and elucidate network structure variables. Our findings provide new insights into understanding the structure of the fast-evolving AI scientific ecosystem and identify the role of influencing factors to acquire central positions in the AI research community. This work may help policymakers and researchers adjust better strategies to make most of the interdisciplinary collaborations and accelerate innovation and knowledge transfer in a highly interdisciplinary field such as AI.

**Limitations and Future Work**

Admittedly, there were some limitations to this study that should be mentioned. First, we used the co-authorship network as a proxy for scientific collaboration. However, scientific collaborative relationships do not necessarily lead to a joint publication and may be resulted from other formal and informal collaboration types. Thus, future research could consider other indicators to measure research collaboration and overcome this limitation. Second, this work identified the most central researchers by calculating common network centrality metrics and categorizing researchers based on their centrality values. Future work may consider other network structural properties or apply other approaches such as unsupervised machine learning techniques to classify core and peripheral researchers and compare the results. Lastly, we considered features that reflected the scientific activities of researchers. Hence, another interesting future research direction would be expanding our feature space and, for example, investigating the effects of casual mechanisms, psychological, and cognitive properties of authors on achieving certain strategic roles in co-authorship networks.